\documentclass[superscriptaddress]{revtex4}
\usepackage{amsfonts}
\usepackage{amsmath}
\usepackage{amssymb}

\input{tcilatex}
\begin{document}

\author{Juan Sebasti\'{a}n Ardenghi}
\affiliation{CONICET-IAFE-Universidad de Buenos Aires, Argentina}
\author{Mario Castagnino}
\affiliation{CONICET-IAFE-IFIR-Universidad de Buenos Aires, Argentina}
\author{Sebastian Fortin}
\affiliation{CONICET-IAFE-Universidad de Buenos Aires, Argentina}
\author{Olimpia Lombardi}
\affiliation{CONICET-Universidad de Buenos Aires, Argentina}

\begin{abstract}
According to our modal-Hamiltonian interpretation (MHI) of quantum
mechanics, the Hamiltonian of the closed system defines the set of its
definite-valued observables. This definition seems to be incompatible with
the pointer basis selected by the environment-induced decoherence (EID) of
the open system. In this paper we argue that decoherence can be understood
from a closed system perspective which (i) shows that the incompatibility
between MHI and EID is only apparent, and (ii) solves certain conceptual
challenges that the EID program still has to face.
\end{abstract}

\title{Decoherence, measurement and interpretation of quantum mechanics}
\maketitle

\section{Introduction}

Since the early days of quantum mechanics, the measurement problem has been
one of the most serious challenges for the interpretation of the theory, and
much ink has been spilled over the search of an adequate solution. During
the last decades, environment-induced decoherence (EID) has become an
unavoidable element in the explanation of quantum measurement (\cite{[1]}, 
\cite{[2]}). The core of the decoherence program relies on the interaction
between the measuring apparatus and its environment: the continuous
\textquotedblleft monitoring\textquotedblright\ of the environment leads
interference to vanish with respect to a definite preferred basis, which
turns out to be the eigenbasis of the pointer observable of the measuring
apparatus. According to Schlosshauer (\cite{[3]}, p.1287; see also \cite%
{[3']}), \textquotedblleft \textit{based on the progress already achieved by
the decoherence program, it is reasonable to anticipate that decoherence
embedded in some additional interpretive structure could lead to a complete
and consistent derivation of the classical world from quantum-mechanical
principles}.\textquotedblright

Recently we have proposed a new interpretation of quantum mechanics (\cite%
{[4]}, \cite{[5]}, \cite{[6]}, \cite{[6']}), belonging to the
\textquotedblleft modal family\textquotedblright : like previous modal
interpretations, our modal-Hamiltonian interpretation (MHI) is a realist,
non-collapse approach, according to which the quantum state describes the
possible properties of the system but not its actual properties (see \cite%
{[6'']}). Then, any modal interpretation is committed to state a rule that
selects the preferred context, that is, the set of the actually
definite-valued observables of the quantum system. According to the MHI, the
preferred context is defined by the Hamiltonian of the system, which is
conceived as a closed system with no external interaction. To the extent
that EID applies to open systems, this seems to mean that the MHI cannot
supply the interpretive structure that would allow decoherence to offer an
adequate account of the classical limit of quantum mechanics. Given the
great success of the EID program, this would count against the acceptability
of the MHI.

The aim of this paper is to show that the conflict between EID and MHI is
merely apparent, and only due to the open-system perspective from which the
theory of decoherence is usually presented. However, decoherence can be
explained from a closed-system perspective, in terms of which the seeming
conflict vanishes. Moreover, from a general viewpoint we will argue that
this closed-system approach solves certain conceptual challenges that the
EID program still has to face.

\section{The measurement problem}

In the standard von Neumann model, a quantum measurement is conceived as an
interaction between a system $S$ and a measuring apparatus $M$. Before the
interaction, $M$\ is prepared in a ready-to-measure state $\left\vert
p_{0}\right\rangle $, eigenvector of the pointer observable $P$\ of $M$, and
the state $\left\vert \psi _{0}\right\rangle $ of $S$\ is a superposition of
the eigenstates $\left\vert a_{i}\right\rangle $ of an observable $A$\ of $S$%
. The interaction introduces a correlation between the eigenstates $%
\left\vert a_{i}\right\rangle $\ of $A$\ and the eigenstates $\left\vert
p_{i}\right\rangle $ of $P$:%
\begin{equation}
\left\vert \psi _{0}\right\rangle =\sum_{i}c_{i}\left\vert
a_{i}\right\rangle \otimes \left\vert p_{0}\right\rangle \longrightarrow
\left\vert \psi \right\rangle =\sum_{i}c_{i}\left\vert a_{i}\right\rangle
\otimes \left\vert p_{i}\right\rangle  \label{1}
\end{equation}%
The problem consists in explaining why, being the state $\left\vert \psi
\right\rangle $\ a superposition of the $\left\vert a_{i}\right\rangle
\otimes \left\vert p_{i}\right\rangle $, the pointer $P$\ acquires a
definite value.

In the orthodox collapse interpretation, the pure state $\left\vert \psi
\right\rangle $ is assumed to \textquotedblleft collapse\textquotedblright\
to a mixture $\rho ^{c}$:%
\begin{equation}
\rho ^{c}=\sum_{i}\left\vert c_{i}\right\vert ^{2}\left\vert
a_{i}\right\rangle \otimes \left\vert p_{i}\right\rangle \left\langle
a_{i}\right\vert \otimes \left\langle p_{i}\right\vert  \label{2}
\end{equation}%
where the probabilities $\left\vert c_{i}\right\vert ^{2}$ are given an
ignorance interpretation. Then, in this situation it is supposed that the
measuring apparatus is in one of the eigenvectors $\left\vert
p_{i}\right\rangle $\ of $P$, say $\left\vert p_{k}\right\rangle $, and
therefore $P$ acquires a definite value $p_{k}$, the eigenvalue
corresponding to the eigenvector $\left\vert p_{k}\right\rangle $, with
probability $\left\vert c_{k}\right\vert ^{2}$. In the modal
interpretations, the problem is to explain the definite reading of the
pointer, with its associated probability, without assuming the collapse
hypothesis.

\section{The MHI account of measurement}

In order to study the physical world, we have to identify the systems that
populate it. By adopting an algebraic perspective, the MHI defines a \textit{%
quantum system} $S$ as a pair $\left( \mathcal{O},H\right) $ such that (i) $%
\mathcal{O}$\ is a space of self-adjoint operators on a Hilbert space $%
\mathcal{H}$, representing the observables of the system, (ii) $H\in 
\mathcal{O}$ is the time-independent Hamiltonian of the system, and (iii) if 
$\rho _{0}\in \mathcal{O}^{^{\prime }}$\ (where $\mathcal{O}^{^{\prime }}$
is the dual space of $\mathcal{O}$) is the initial state of $S$, $\rho _{0}$
evolves according to the Schr\"{o}dinger equation in its von Neumann version.

Of course, any quantum system can be decomposed in parts in many ways;
however, not any decomposition will lead to parts which are, in turn,
quantum systems. According to the MHI, a quantum system $S:\left( \mathcal{O}%
,H\right) $\ with initial state $\rho _{0}\in \mathcal{O}^{^{\prime }}$ is 
\textit{composite} when it can be partitioned into two quantum systems $%
S_{1}:\left( \mathcal{O}_{1},H_{1}\right) $ and $S_{2}:\left( \mathcal{O}%
_{2},H_{2}\right) $ such that (i) $\mathcal{O}=\mathcal{O}_{1}\otimes 
\mathcal{O}_{2}$, and (ii) $H=H_{1}\otimes I_{2}+I_{1}\otimes H_{2}$\ (where 
$I_{1}$ and $I_{2}$ are the identity operators in the corresponding tensor
product spaces). In this case, the initial states of $S_{1}$ and $S_{2}$ are
obtained as the partial traces $\rho _{01}=Tr_{(2)}\rho _{0}$ and $\rho
_{02}=Tr_{(1)}\rho _{0}$, and we say that $S_{1}$ and $S_{2}$ are \textit{%
subsystems} of the composite system $S$.

If the quantum system is not composite, we call it \textit{elemental}. There
are different, equally legitimate ways of decomposing an elemental system $%
S:\left( \mathcal{O},H\right) $ with initial state $\rho _{0}\in \mathcal{O}%
^{^{\prime }}$ into \textquotedblleft parts\textquotedblright\ $P_{1}:\left( 
\mathcal{O}_{1},H_{1}\right) $ and $P_{2}:\left( \mathcal{O}%
_{2},H_{2}\right) $, with initial states $\rho _{01}\in \mathcal{O}%
_{1}^{^{\prime }}$ and $\rho _{02}\in \mathcal{O}_{2}^{^{\prime }}$
respectively, such that (i) $\mathcal{O}=\mathcal{O}_{1}\otimes \mathcal{O}%
_{2}$, (ii) $H=H_{1}\otimes I_{2}+I_{1}\otimes H_{2}+H_{12}^{int}$, where $%
H_{12}^{int}$ is the interaction Hamiltonian, and (iii) $\rho
_{01}=Tr_{(2)}\rho _{0}$ and $\rho _{02}=Tr_{(1)}\rho _{0}$ \ (see \cite{[7]}%
, \cite{[8]}). Since $H_{12}^{int}\neq 0$, $\rho _{01}$ and $\rho _{02}$ do
not evolve unitarily according to the Schr\"{o}dinger equation: it is for
this reason that $P_{1}$ and $P_{2}$ are not subsystems but have to be
considered as mere \textit{parts} of $S$.

Given the contextuality of quantum mechanics, the subtler point in any
realist interpretation is the selection of the preferred context. In the MHI
this selection is based on the \textit{actualization rule}, which defines,
among all the observables of the system, those that acquire actual, and not
merely possible definite values. The basic idea is that the Hamiltonian of
the system defines actualization; therefore, any observable that does not
have the symmetries of the Hamiltonian cannot acquire an actual value, since
its actualization would break the symmetry of the system in an arbitrary
way. Precisely, the MHI actualization rule states that, given an elemental
quantum system $S:\left( \mathcal{O},H\right) $, the \textit{preferred
context} consists of $H$ and the observables commuting with $H$ and having,
at least, the same symmetries as $H$. This rule has been applied to many
well-known physical situations (hydrogen atom, Zeeman effect, fine
structure, etc.), leading to results consistent with experimental evidence
(see \cite{[4]}, \cite{[5]}). Moreover, it has proved to be effective for
solving the measurement problem, both in its ideal and its non-ideal
versions; for our purposes, we will only focalize on measurement.

According to the MHI, a quantum measurement is a three-stage process. In the
first stage, the system $S$ to be measured $-$represented in the Hilbert
space $\mathcal{H}_{S}$ and with Hamiltonian $H_{S}-$ and the measuring
device $D$ $-$represented in the Hilbert space $\mathcal{H}_{D}$ and with
Hamiltonian $H_{D}-$ do not interact. During the second stage, an
interaction Hamiltonian $H_{SD}^{int}$ introduces the correlation between
the eigenstates $\left\vert a_{i}\right\rangle $ of an observable $A$ of $S$
and the eigenstates $\left\vert p_{i}\right\rangle $ of a pointer observable 
$P$ of $D$ (see \cite{[9]}). In the third stage the interaction ends, and
the whole system becomes a composite system $S\cup D$ with a Hamiltonian%
\begin{equation}
H=H_{S}\otimes I_{D}+I_{S}\otimes H_{D}  \label{3}
\end{equation}%
and an initial state (see eq.(\ref{1}))%
\begin{equation}
\left\vert \psi _{SD}\right\rangle =\sum_{i}c_{i}\left\vert
a_{i}\right\rangle \otimes \left\vert p_{i}\right\rangle  \label{4}
\end{equation}%
Although $\left\vert \psi _{SD}\right\rangle $ is an entangled state, since
there is no interaction between the subsystems $S$ and $D$, the
actualization rule has to be applied to each one of them independently. In
particular, when applied to $D$, the rule states that the definite-valued
observables are the Hamiltonian $H_{D}$ and all the observables commuting
with $H_{D}$ and having, at least, the same symmetries $-$degeneracies$-$ as 
$H_{D}$.

Of course, not any quantum process can be considered a quantum measurement.
On the basis of the above description, we can formulate the conditions that
define a quantum measurement:

\begin{enumerate}
\item[(a)] During a period $\Delta t$, $S$ and $D$ must interact through an
interaction Hamiltonian $H_{SD}^{int}\neq 0$ intended to introduce a
correlation between the observable $A$ of $S$ and the pointer $P$ of $D$.
The requirement of perfect correlation is not included as a defining
condition of measurement, because the actualization rule explains the
definite reading of the pointer $P$ even in non-ideal measurements, that is,
when the correlation is not perfect. In this case, the rule also accounts
for the difference between \textit{reliable} and \textit{non}-\textit{%
reliable} measurements (see \cite{[5]}, Section 6).

\item[(b)] The measuring device $D$ has to be such that its pointer $P$ (i)
has macroscopically distinguishable eigenvalues, and (ii) commutes with the
Hamiltonian $H_{D}$ and has, at least, the same degeneracy as $H_{D}$. The
condition $\left[ P,H_{D}\right] =0$, besides to explain the definite
reading of the pointer according to the MHI actualization rule, guarantees
the stationarity of the eigenvectors of $P$, making the readings of the
pointer possible.
\end{enumerate}

This account of the quantum measurement has been used to explain how the
initial $-$pure or mixed$-$ state is reconstructed through measurement both
in the ideal and in the non-ideal case, and has been successfully applied to
the paradigmatic example of the Stern-Gerlach experiment, with perfect and
non perfect correlation, and also in the case of an imperfect collimation of
the incoming beam (see \cite{[5]}, Section 6).

\section{The EID account of measurement}

As it is well-known, the key idea of the decoherence program is that
macroscopic systems, like measuring apparatuses, are never isolated but
always interact with their environments. In the von Neumann model of
measurement (see eq.(\ref{1})), when the environment $E$ is taken into
account, after the correlation the initial state of the whole system $S+M+E$
becomes%
\begin{equation}
\left\vert \psi _{SME}(0)\right\rangle =\left( \sum_{i}c_{i}\left\vert
a_{i}\right\rangle \otimes \left\vert p_{i}\right\rangle \right) \otimes
\left\vert e_{0}\right\rangle   \label{5}
\end{equation}%
where $\left\vert e_{0}\right\rangle $ is the state of the environment
before its interaction with the measuring apparatus. Zurek and his
collaborators prove that, when the interaction Hamiltonian between the
measuring apparatus $M$ and the environment $E$, $H_{ME}^{int}$, satisfies
certain conditions (see \cite{[10]}), $\left\vert \psi
_{SME}(0)\right\rangle $ evolves into%
\begin{equation}
\left\vert \psi _{SME}(t)\right\rangle =\sum_{i}c_{i}\left\vert
a_{i}\right\rangle \otimes \left\vert p_{i}\right\rangle \otimes \left\vert
e_{i}(t)\right\rangle   \label{6}
\end{equation}%
where the $\left\vert e_{i}(t)\right\rangle $ are the states of the
environment associated with the different pointer states $\left\vert
p_{i}\right\rangle $. According to Zurek, the state of the system $S+M$ is
represented by the reduced density operator $\rho _{r}(t)$ resulting from
tracing over the environmental degrees of freedom,%
\begin{eqnarray}
\rho _{r}(t) &=&Tr_{(E)}\left( \left\vert \psi _{SME}(t)\right\rangle
\left\langle \psi _{SME}(t)\right\vert \right)   \notag \\
&=&\sum_{ij}c_{i}c_{j}^{\ast }\left\vert a_{i}\right\rangle \otimes
\left\vert p_{i}\right\rangle \left\langle a_{j}\right\vert \otimes
\left\langle p_{j}\right\vert \sum_{l}\left\langle
e_{l}(t)|e_{i}(t)\right\rangle \left\langle e_{j}(t)|e_{l}(t)\right\rangle 
\label{7}
\end{eqnarray}%
where the factor $r_{ij}(t)=\sum_{l}\left\langle
e_{l}(t)|e_{i}(t)\right\rangle \left\langle e_{j}(t)|e_{l}(t)\right\rangle $
determines the size of the off-diagonal terms at each time. Many standard
models for the interaction Hamiltonian $H_{ME}^{int}$ show that, when the
environment is composed of a large number of subsystems, the states $%
\left\vert e_{i}(t)\right\rangle $ of the environment rapidly approach
orthogonality. This means that the reduced density operator rapidly becomes
approximately diagonal in the preferred basis $\left\{ \left\vert
a_{i}\right\rangle \otimes \left\vert p_{i}\right\rangle \right\} $ (compare
with eq.(\ref{2})),%
\begin{equation}
r_{ij}(t)\rightarrow \delta _{ij}\quad \Rightarrow \quad \rho
_{r}(t)\rightarrow \rho _{r}=\sum_{i}\left\vert c_{i}\right\vert
^{2}\left\vert a_{i}\right\rangle \otimes \left\vert p_{i}\right\rangle
\left\langle a_{i}\right\vert \otimes \left\langle p_{i}\right\vert 
\label{8}
\end{equation}%
According to Zurek (\cite{[10]}, \cite{[1]}), in a certain sense decoherence
\textquotedblleft explains\textquotedblright\ the collapse of the state
vector.

In his first papers, Zurek studied physical models where the interaction
between the measuring apparatus and the environment dominates the process (%
\cite{[10]}, \cite{[11]}); in those cases, the reduced density matrix ends
up being diagonal in the eigenvectors of an observable $P$ that commutes
with the Hamiltonian $H_{ME}^{int}$ describing the apparatus-environment
interaction. This property is what makes $P$ to be the pointer observable:
since $P$ is a constant of motion of $H_{ME}^{int}$, when the apparatus is
in one of its eigenstates, the interaction with the environment will leave
it unperturbed. Since those first works, the condition $\left[ P,H_{ME}^{int}%
\right] =0$ has usually be considered as the definition of the pointer basis
or of the pointer observable $P$ of the apparatus (see, for instance, \cite%
{[12]}, p.363, \cite{[3]}, pp.1278-1279).

In the 90's, Zurek stressed that the original definition of the pointer
basis was a simplification: in more general situations, when the system's
dynamics is relevant, the einselection of the preferred basis is more
complicated. Zurek introduced the \textquotedblleft \textit{predictability
sieve}\textquotedblright\ criterion (\cite{[13]}, \cite{[14]}) as a
systematic strategy to identify the preferred basis in generic situations.
The criterion is based on the fact that the preferred states are, by
definition, those less affected by the interaction with the environment, in
the sense that they are the ones less entangled with it. On the basis of the
application of this criterion, three basically different regimes for the
selection of the preferred basis can be distinguished (\cite{[15]}, \cite%
{[2]}):

\begin{itemize}
\item The first regime is the quantum measurement situation, where the
self-Hamiltonian of the system can be neglected and the evolution is
completely dominated by the interaction Hamiltonian. In such a case, the
preferred states are directly the eigenstates of the interaction Hamiltonian
(\cite{[10]}).

\item The second regime is the more realistic and complex situation, where
neither the self-Hamiltonian of the system nor the interaction with the
environment are clearly dominant, but both induce non-trivial evolution. In
this case, the preferred basis arises from the interplay between
self-evolution and interaction; quantum Brownian motion belongs to this case
(\cite{[16]}).

\item The third regime corresponds to the situation where the dynamics is
dominated by the system's self-Hamiltonian. In this case, the preferred
states are simply the eigenstates of this self-Hamiltonian (\cite{[15]}).
\end{itemize}

According to Schlosshauer (\cite{[3]}, p.1280), these three regimes explain
why many systems, specially in the macroscopic domain, are typically found
in energy eigenstates, even if the interaction Hamiltonian depends on an
observable different than energy.

\section{Measurement from a closed-system perspective}

As we have seen, the actualization rule of the MHI explains the definite
reading of the pointer $P$ of the measuring device $D$ by considering that $P
$ commutes with the Hamiltonian $H_{D}$ of $D$ and does not break the
degeneracies of such a Hamiltonian. This account of the quantum measurement
seems to be at odds with the explanation given by the EID program, according
to which the decoherence of the measuring apparatus in interaction with its
environment is what causes the apparent \textquotedblleft
collapse\textquotedblright\ that suppresses superpositions. In fact, in the
MHI, the environment is absent: after the interaction $D$ is a closed
quantum system unitarily evolving with its own Hamiltonian $H_{D}$.
Moreover, this seems to flagrantly contradict the fact that real measuring
apparatuses are never isolated, but they interact significantly with their
environments. However, this apparent conflict vanishes when a
\textquotedblleft closed-system\textquotedblright\ perspective is adopted.

\subsection{Revisiting the MHI account of measurement}

If measurement is described in terms of the quantum and, therefore, closed
systems involved in the process, the measuring device $D$ has to be
considered \textit{not as an open macroscopic apparatus} $A$ (eventually
surrounded by a \textquotedblleft bath\textquotedblright\ $B$ of particles
in interaction with it), but as the entire quantum system that interacts
with the system $S$ in the second stage and remains closed in the third
stage: it is this system what has to have a pointer observable commuting
with its Hamiltonian $H_{D}$. On this basis, we can now analyze the elements
participating in the process as described in the framework of the MHI:

\begin{itemize}
\item The closed system $D$ $-$e.g., the open macroscopic apparatus $A$ 
\textit{plus} the bath of particles $B-$ is certainly a macroscopic system,
whose Hamiltonian is the result of the interaction among a huge number of
degrees of freedom. Since, in general, symmetries are broken by
interactions, the symmetry of a Hamiltonian decreases with the complexity of
the system. Then, a macroscopic system having a Hamiltonian with symmetries
is a highly exceptional situation: in the generic case, the energy is the
only constant of motion of the macroscopic system. As a consequence, in
realistic measurement situations, $H_{D}$ is non-degenerate,%
\begin{equation}
H_{D}\left\vert \omega _{k}\right\rangle =\omega _{k}\left\vert \omega
_{k}\right\rangle \text{ with }\omega _{k}\neq \omega _{k^{\prime }}
\label{9}
\end{equation}%
and, therefore, $\left\{ \left\vert \omega _{k}\right\rangle \right\} $ is a
basis of the Hilbert space $\mathcal{H}_{D}$ of $D$. This means that, when $%
\left[ P,H_{D}\right] =0$ , we can guarantee that $P$ has, at least, the
same degeneracies as $H_{D}$ because $H_{D}$ is non-degenerate.

\item The pointer $P$ cannot have such a huge number of different
eigenvalues as $H_{D}$, because the experimental physicist must be able to
discriminate among them (for instance, in the Stern-Gerlach experiment the
pointer has three eigenvalues). This means that $P$ is a \textquotedblleft
collective\textquotedblright\ observable of the closed system $D$ (see \cite%
{[17]}, \cite{[18]}), that is, a highly degenerate observable that does not
\textquotedblleft see\textquotedblright\ the vast majority of the degrees of
freedom of $D$:%
\begin{equation}
P=\sum_{n}p_{n}P_{n}  \label{10}
\end{equation}%
where the set $\left\{ P_{n}\right\} $ of the eigenprojectors of $P$ spans
the Hilbert space $\mathcal{H}_{D}$ of $D$. If we call $N$ the number of
eigenprojectors of $P$, and $K$ the dimension of $\mathcal{H}_{D}$, then it
is clear that $K\gg N$. In other words, the eigenprojectors of $P$ introduce
a sort of \textquotedblleft coarse-graining\textquotedblright\ onto the
Hilbert space $\mathcal{H}_{D}$. Therefore, if the Hamiltonian $H_{D}$ is
non-degenerate, the condition $\left[ P,H_{D}\right] =0$ implies that $P$
can be expressed in terms of the energy eigenbasis $\left\{ \left\vert
\omega _{k}\right\rangle \right\} $ as%
\begin{equation}
P=\sum_{n}p_{n}P_{n}=\sum_{n}p_{n}\sum_{i_{n}}\left\vert \omega
_{i_{n}}\right\rangle \left\langle \omega _{i_{n}}\right\vert   \label{11}
\end{equation}%
This expression shows that, since $p_{n}\neq p_{n^{\prime }}$, $P$ has much
more degeneracies than $H_{D}$.

\item The requirement $\left[ P,H_{D}\right] =0$, far from being an \textit{%
ad hoc} condition necessary to apply the MHI actualization rule, has a clear
physical meaning: it is essential to preserve the stationary behavior of $P$
during the third stage of the measurement process. If this requirement did
not hold because of the uncontrollable interaction among the microscopic
degrees of freedom of the macroscopic device or between the macroscopic
device and an external \textquotedblleft bath\textquotedblright , the
reading of $P$ would constantly change and measurement would be impossible.
This goal may be achieved by many different technological means; but, in any
case, measurement has to be a controlled situation where the reading of a
stable pointer can be obtained.
\end{itemize}

\subsection{The pointer basis from a closed-system perspective}

In the context of EID, during the third stage the measuring apparatus $M$
does no longer interact with the measured system $S$ but interacts with the
environment $E$. If we call, as before, $D=M+E$ the whole system that
interacts with $S$ in the second stage but remains closed during the third
stage, the question is how to identify the open parts of $D$ to be conceived
as the measuring apparatus $M$ and as the environment $E$. This is a
legitimate question because, as we have pointed out, a whole closed system
may be partitioned in many different ways, none of them more
\textquotedblleft essential\textquotedblright\ than the others (see \cite%
{[7]}, \cite{[8]}).

A natural assumption is to consider the macroscopic, material apparatus $A$
built for measurement as \textquotedblleft the measuring
apparatus\textquotedblright\ $M$, and the bath $B$ of the particles
scattering off $A$ as \textquotedblleft the environment\textquotedblright\ $%
E $; then, $D=A+B$ is the closed system resulting from the interaction
between $A$ and B. From this position, it is supposed that $A$ is the open
system that decoheres: the reduced density operator $\rho _{r}^{A}(t)$ of $A$
should converge to a final time-independent $\rho _{r}^{A}$, diagonal in the
preferred basis of $A$, that is, of its Hilbert space $\mathcal{H}_{A}$, and
the pointer $P$ should define such a basis. However, even if apparently
\textquotedblleft natural\textquotedblright , this is not the best choice
for the splitting of $D$, since it does not take into account the
environment \textit{internal} to the device $A$. In fact, being a
macroscopic body, $A$ also has a huge number of degrees of freedom, which
have to be \textquotedblleft coarse-grained\textquotedblright\ by $P$ if it
is to play the role of the pointer. In other words, since the pointer $P$
must have a small number of different eigenvalues to allow the experimenter
to discriminate among them, $P$ is a highly degenerate observable on the
Hilbert space $\mathcal{H}_{A}$ of the open macroscopic apparatus $A$ and,
as a consequence, it does not define a \textit{basis} of $\mathcal{H}_{A}$.

When we recall that the only univocally definable entity is the $-$closed$-$
quantum system, and that a quantum system can be partitioned in many,
equally legitimate manners, the closed system $D$ can be split in a
theoretically best founded way in the measurement case. Let us recall that
the pointer $P$ is the observable whose eigenvectors became correlated with
the eigenvectors of an observable of the measured system during the second
stage of the process, and that the interaction in that stage was
deliberately designed to introduce such a correlation. So, if we want that
during the third stage $P$ really defines a \textit{basis}, the open
\textquotedblleft measuring apparatus\textquotedblright\ $M$ must be the
part of $D$ corresponding to the Hilbert space $\mathcal{H}_{M}$ where the
pointer is non-degenerate. If we call $P_{M}$ the pointer belonging to $%
\mathcal{H}_{M}\otimes \mathcal{H}_{M}$, it reads%
\begin{equation}
P_{M}=\sum_{n}p_{n}\left\vert p_{n}\right\rangle \left\langle
p_{n}\right\vert   \label{12}
\end{equation}%
where $\left\{ \left\vert p_{n}\right\rangle \right\} $ is a basis of $%
\mathcal{H}_{M}$. Then, the relevant partition is $\mathcal{H}_{D}=\mathcal{H%
}_{M}\otimes \mathcal{H}_{E}$, where $\mathcal{H}_{E}$ is the Hilbert space
of the \textquotedblleft environment\textquotedblright\ $E$. If $\left\{
\left\vert e_{n}\right\rangle \right\} $ is a basis of $\mathcal{H}_{E}$,
the pointer acting on $\mathcal{H}_{D}$ can be expressed as a highly
degenerate observable:%
\begin{eqnarray}
P &=&P_{M}\otimes I_{E}=\left( \sum_{n}p_{n}\left\vert p_{n}\right\rangle
\left\langle p_{n}\right\vert \right) \otimes \left( \sum_{m}\left\vert
e_{m}\right\rangle \left\langle e_{m}\right\vert \right)   \notag \\
&=&\sum_{n}p_{n}\sum_{m}\left\vert p_{n}\right\rangle \otimes \left\vert
e_{m}\right\rangle \left\langle p_{n}\right\vert \otimes \left\langle
e_{m}\right\vert =\sum_{n}p_{n}P_{n}  \label{13}
\end{eqnarray}%
This agrees with the features of $P$ in the MHI: $P$ introduces a sort of
\textquotedblleft coarse-graining\textquotedblright\ onto the Hilbert space $%
\mathcal{H}_{D}$ (compare eq.(\ref{13}) with eq.(\ref{10})). The many
degrees of freedom corresponding to the degeneracies of $P$ in $\mathcal{H}%
_{D}$ play the role of the \textquotedblleft environment\textquotedblright\ $%
E$, composed by the microscopic degrees of freedom of the macroscopic
apparatus $A$ $-$internal environment$-$ and the degrees of freedom of the
bath $B$ $-$external environment$-$.

\subsection{The agreement between EID and MHI}

As we have seen, in the first papers on decoherence, the condition $\left[
P,H_{ME}^{int}\right] =0$ was considered as the definition of the pointer
basis. However, this definition involves several assumptions. In fact, from
a closed-system perspective, the entangled state $\left\vert \psi
_{SME}(t)\right\rangle $ of the whole system actually evolves according to
the Schr\"{o}dinger equation under the action of the total Hamiltonian $%
H_{SME}=H_{S}+H_{M}+H_{E}+H_{SM}^{int}+H_{SE}^{int}+H_{ME}^{int}$. So, first
it is considered that the system-environment interaction and the
system-apparatus interaction are zero: $H_{SE}^{int}=0$ and $H_{SM}^{int}=0$%
. This assumption is reasonable on the basis of the design of the
measurement arrangement: after a short time, any interaction with the system
ends and the subsystem $M+E$ follows its independent dynamical evolution;
for this reason, also the self-Hamiltonian $H_{S}$ of the system can be
disregarded. Then, the stability of the pointer strictly requires that 
\begin{equation}
\left[ P,H_{ME}\right] =0  \label{14}
\end{equation}%
where $H_{ME}$, when expressed in precise terms, reads%
\begin{equation}
H_{ME}=H_{M}\otimes I_{E}+I_{M}\otimes H_{E}+H_{ME}^{int}  \label{15}
\end{equation}%
If we recall that the pointer is an observable $P$ highly degenerate in the $%
-$internal and external$-$ degrees of freedom of the environment (see eq.(%
\ref{13})), condition (\ref{14}) results%
\begin{equation}
\left[ P,H_{ME}\right] =\left[ P_{M}\otimes I_{E},H_{M}\otimes
I_{E}+I_{M}\otimes H_{E}+H_{ME}^{int}\right] =0  \label{16}
\end{equation}%
But since always $\left[ P_{M}\otimes I_{E},I_{M}\otimes H_{E}\right] =0$,
then the stability requirement for the pointer observable becomes that it
commute with the Hamiltonian $H_{M}\otimes I_{E}+H_{ME}^{int}$, where the
self-Hamiltonian of the environment is not included:%
\begin{equation}
\left[ P,H_{M}\otimes I_{E}+H_{ME}^{int}\right] =0  \label{17}
\end{equation}

This argument shows that the condition $\left[ P,H_{ME}^{int}\right] =0$
introduced in the first papers on decoherence, that is, that the pointer
commutes with the interaction Hamiltonian, is a particular case which holds
only when the self-Hamiltonian of $M$ can be disregarded. It is also clear
that the three regimes, distinguished by Zurek as the result of the
application of the predictability sieve to a number of models (see Section
IV), turn out to be the three particular cases of condition (\ref{17}), and
can be redescribed in terms of that condition:

\begin{itemize}
\item When $H_{M}\otimes I_{E}\ll H_{ME}^{int}$, the self-Hamiltonian of $M$
can be neglected, and then $\left[ P,H_{ME}^{int}\right] =0$. Therefore, the
preferred basis is defined by the interaction Hamiltonian $H_{ME}^{int}$.

\item When $H_{M}\otimes I_{E}\cong H_{ME}^{int}$, neither the
self-Hamiltonian of $M$ nor the interaction with the environment are clearly
dominant. In this case, the preferred basis is defined by condition (\ref{17}%
).

\item When $H_{M}\otimes I_{E}\gg H_{ME}^{int}$, the dynamics is dominated
by self-Hamiltonian of $M$ and, then, $\left[ P,H_{M}\otimes I_{E}\right] =%
\left[ P_{M}\otimes I_{E},H_{M}\otimes I_{E}\right] =\left[ P_{M},H_{M}%
\right] =0$. Therefore, the preferred states are simply the eigenstates of $%
H_{M}$.
\end{itemize}

As a consequence, the fact, noted by Schlosshauer (\cite{[3]}, p.1280), that
many systems are typically found in energy eigenstates although the
interaction Hamiltonian depends on an observable different than energy, far
from being surprising, necessarily results from the requirement of stability
for the preferred basis. But the point we want to stress here is that, when
the EID pointer basis is considered from this closed-system viewpoint, it
agrees with the preferred context as defined by the MHI actualization rule:
in both cases, the pointer/preferred basis is given by the Hamiltonian of
the whole closed system. In fact, the three regimes identified by Zurek and
obtained case by case by means of the predictability sieve criterion, turn
out to be particular cases of the MHI characterization of the preferred
basis: if the preferred states are defined by the eigenstates of the
Hamiltonian of the whole system, it is not hard to realize that they will
depend on the Hamiltonian's component which dominates the whole evolution.

Moreover, when the pointer basis is viewed from this closed-system
perspective, Zurek's first regime can be justified on general grounds.
According to Zurek, the first regime is the quantum measurement situation,
where the self-Hamiltonian of the measuring system $M$ can be neglected and
the evolution is completely dominated by the interaction Hamiltonian: this
means that $H_{M}\otimes I_{E}\ll H_{ME}^{int}$. If the apparatus is now
conceived as the part of the closed system $D$ \textquotedblleft
viewed\textquotedblright\ by the pointer $P$, and the environment carries
over almost all the degrees of freedom of $D$, it seems reasonable to
suppose that, in general, the Hamiltonian corresponding to the interaction
with that huge number of degrees of freedom is much greater than the
self-Hamiltonian of the \textquotedblleft small\textquotedblright\ part
defined by the pointer. Therefore, the condition $H_{M}\otimes I_{E}\ll
H_{ME}^{int}$ leading to the first regime turns out to have a physical
justification.

\section{EID from a closed-system perspective}

Zurek considers that the prejudice which seriously delayed the solution of
the problem of the transition from quantum to classical is itself rooted in
the fact that the role of the \textquotedblleft openness\textquotedblright\
of a quantum system in the emergence of classicality was ignored for a very
long time (\cite{[1]}, \cite{[2]}). However, since the environment may be
external or internal, the decoherence program supplies no general criterion
for distinguishing between the system and its environment: the partition of
the whole closed system is decided case by case, and usually depends on the
previous assumption of the observables that will behave classically (for a
discussion of this point, see \cite{[19]}). Zurek recognizes this problem as
a shortcoming of his proposal: \textquotedblleft \textit{In particular, one
issue which has been often taken for granted is looming big as a foundation
of the whole decoherence program. It is the question of what are the
`systems' which play such a crucial role in all the discussions of the
emergent classicality. This issue was raised earlier, but the progress to
date has been slow at best}\textquotedblright\ (\cite{[20]}, p.22).

Of course, the problem of defining the systems involved in decoherence seems
to be a serious obstacle for the einselection program when the phenomenon is
described in the usual terms, that is, as a consequence of the interaction
between two open systems. However, as emphasized by Omn\`{e}s (\cite{[21]}, 
\cite{[22]}), decoherence can be conceived as a particular case of the
general phenomenon of irreversibility, where the non-unitary evolution is
obtained by disregarding part of the maximal information obtainable from the
system. In the quantum case, the maximal information about a system is given
by the set $\mathcal{O}$ of all its possible observables; then, the
restriction of that maximal information to a relevant part amounts to select
a subset $\mathcal{O}^{R}\subset \mathcal{O}$\ of relevant observables, with
respect to which the behavior of the system will be studied. From this
perspective, the identification of the system of interest $S$ and its
environment $E$ in the decoherence program amounts to a particular selection
of the relevant observables of the whole closed system. In particular, if a
closed quantum system $U$ is represented in the Hilbert space $\mathcal{H}$
and its observables $O$ belong to the von Neumann-Liouville space $\mathcal{L%
}=\mathcal{H}\otimes \mathcal{H}$, the relevant observables $O^{R}$ are
those of the form 
\begin{equation}
O^{R}=O_{S}\otimes I_{E}\in \mathcal{O}^{R}\subset \mathcal{L}  \label{18}
\end{equation}%
where the $O_{S}\in \mathcal{L}_{S}=\mathcal{H}_{S}\otimes \mathcal{H}_{S}$
are the observables of $S$, $I_{E}$ is the identity operator on the von
Neumann-Liouville space $\mathcal{L}_{E}=\mathcal{H}_{E}\otimes \mathcal{H}%
_{E}$ of $E$, and $\mathcal{L}=\mathcal{L}_{S}\otimes \mathcal{L}_{E}$. This
means that, since the system $S$ is characterized by its von
Neumann-Liouville space $\mathcal{L}_{S}$ (or, equivalently, by its Hilbert
space $\mathcal{H}_{S}$), when the relevant observables $O^{R}$ are
selected, the system $S$ and, with it, the environment $E$ turn out to be
precisely identified. In turn, by definition of reduced density operator, $%
\rho _{r}$ of $S$ is such that (see \cite{[23]})%
\begin{equation}
\left\langle O^{R}\right\rangle _{\rho }=\left\langle O_{S}\right\rangle
_{\rho _{r}}  \label{19}
\end{equation}%
Therefore, the convergence of $\rho _{r}(t)$ to a final $\rho _{r}$ diagonal
in the preferred basis means that the expectation values $\left\langle
O^{R}\right\rangle _{\rho (t)}=\left\langle O_{S}\right\rangle _{\rho
_{r(t)}}$ approach final stable values: 
\begin{equation}
\rho _{r(t)}\rightarrow \rho _{r}\quad \Rightarrow \quad \left\langle
O^{R}\right\rangle _{\rho (t)}=\left\langle O_{S}\right\rangle _{\rho
_{r(t)}}\rightarrow \left\langle O_{S}\right\rangle _{\rho _{r}}  \label{20}
\end{equation}%
In other words, as in the general case of irreversibility, in the process of
decoherence the non-unitary evolution of the reduced state expresses the
convergence of the expectation values to their final values, for all the
observables selected as relevant in each particular case (for a detailed
discussion, see \cite{[24]}).

On the basis of these ideas, in previous papers we have proposed a general,
closed-system theoretical framework for decoherence, where also the EID
approach finds its place (\cite{[24]}, \cite{[25]}); from this perspective
we have computed decoherence times \cite{[26]} and have treated a
generalization of the spin-bath model (\cite{[27]}). But the point to stress
here is that, in this closed-system context, the identification of the
system of interest and its environment is just a way of selecting the
relevant observables of the whole closed system. Since there are many
different sets of relevant observables depending on the observational
viewpoint adopted, the same closed system can be decomposed in many
different ways: each decomposition represents a decision about which degrees
of freedom are relevant and which can be disregarded in any case. Since
there is no privileged or \textquotedblleft essential\textquotedblright\
decomposition, there is no need of an unequivocal criterion to decide where
to place the cut between \textquotedblleft the\textquotedblright\ system and
\textquotedblleft the\textquotedblright\ environment. Therefore, the
\textquotedblleft looming big\textquotedblright\ problem of defining the
systems involved in decoherence is not as serious as Zurek himself supposes:
decoherence is relative to the relevant observables selected in each
particular case.

\section{Conclusions}

At present it is quite clear that the theory of decoherence does not supply
an interpretation of quantum mechanics. Nevertheless, given its impressive
empirical success, it is also clear that nowadays no interpretation can
ignore the results coming from the EID approach. Our MHI has been
successfully applied to many well-known physical situations, and has proved
to be effective for solving the measurement problem, both in its ideal and
its non-ideal versions. However, since the actualization rule applies to
closed systems, the MHI seems to stand at odds of EID.

In this paper we have shown that this assumption is misguided. On the
contrary, when the measurement process is viewed from a closed-system
perspective, the MHI and the EID accounts of measurement agree: the
classical-like states einselected by the interaction with the environment
(the eigenvectors of the pointer, elements of the pointer basis) are the
eigenvectors of an actual-valued observable belonging to the preferred
context according to the MHI.

Moreover, we have argued that, ironically, the \textquotedblleft looming
big\textquotedblright\ problem of defining the systems involved in
decoherence is the consequence of what has been considered to be the main
advantage of the decoherence program: its \textquotedblleft
open-system\textquotedblright\ perspective. Therefore, the closed-system
approach also solves this seeming difficulty, by showing that decoherence is
a relative phenomenon that depends on the partition of the closed system
that is selected in each particular case.\bigskip 

{\Large Acknowledgements}

We are very grateful to Rodolfo Gambini for his kind hospitality at the
Universidad de la Rep\'{u}blica in Montevideo, and for his stimulating
comments on this subject. This work has been supported by grants of CONICET,
ANPCyT, UBA and SADAF, Argentina.

\end{document}